\documentclass[12pt]{article}
\pdfoutput=1
\usepackage{graphicx}
\usepackage{amsmath}
\usepackage{amssymb}
\usepackage{amsfonts}
\usepackage{tensor}
\usepackage{subcaption}
\usepackage[usenames,dvipsnames]{xcolor}
\usepackage[normalem]{ulem}
\usepackage[bottom]{footmisc}
\usepackage{hyperref}

\numberwithin{equation}{section}

\definecolor{orange}{rgb}{1,0.5,0}
\definecolor{grey}{rgb}{.5,.5,.5}
\definecolor{bluegreen}{rgb}{0,.5,.5}
\definecolor{darkgreen}{rgb}{0,.5,0}

\def\gsim{\, \rlap{$>$}{\lower 1.1ex\hbox{$\sim$}}\,}
\def\lsim{\, \rlap{$<$}{\lower 1.1ex\hbox{$\sim$}}\,}
\newcommand{\be}{\begin{equation}}
\newcommand{\ee}{\end{equation}}
\newcommand{\bea}{\begin{eqnarray}}
\newcommand{\eea}{\end{eqnarray}}

\begin{document}


\begin{titlepage}
\bigskip
\bigskip\bigskip\bigskip
\centerline{\Large \bf Combing gravitational hair in 2+1 dimensions}

\bigskip\bigskip\bigskip
\bigskip\bigskip\bigskip

 \centerline{{\bf William Donnelly,}\footnote{\tt donnelly@physics.ucsb.edu}
 {\bf Donald Marolf,}\footnote{\tt marolf@physics.ucsb.edu}
 {and \bf Eric Mintun}\footnote{\tt mintun@physics.ucsb.edu}}
 \bigskip
\centerline{\em Department of Physics}
\centerline{\em University of California}
\centerline{\em Santa Barbara, CA 93106 USA}

\bigskip\bigskip\bigskip

\begin{abstract}
The gravitational Gauss law requires any addition of energy to be accompanied by the addition of gravitational flux. The possible configurations of this flux for a given source may be called gravitational hair, and several recent works discuss gravitational observables (`gravitational Wilson lines') which create this hair in highly-collimated `combed' configurations.  We construct and analyze time-symmetric classical solutions of 2+1 Einstein-Hilbert gravity such as might be created by smeared versions of such operators.  We focus on the AdS$_3$ case, where this hair is characterized by the profile of the boundary stress tensor; the desired solutions are those where the boundary stress tensor at initial time $t=0$ agrees precisely with its vacuum value outside an angular interval $[-\alpha,\alpha]$.    At linear order in source strength the energy is independent of the combing parameter $\alpha$, but non-linearities cause the full energy to diverge as $\alpha \to 0$.  In general, solutions with combed gravitational flux also suffer from what we call displacement from their naive location.  For weak sources and large $\alpha$ one may set the displacement to zero by further increasing the energy, though for strong sources and small $\alpha$ we find no preferred notion of a zero-displacement solution. In the latter case we conclude that naively-expected gravitational Wilson lines do not exist.  In the zero-displacement case, taking the AdS scale $\ell$ to infinity gives finite-energy flux-directed solutions that may be called asymptotically flat.
\end{abstract}

\end{titlepage}

\setcounter{footnote}{0}

\tableofcontents

\section{Introduction}

The AdS/CFT correspondence provides a framework to address many questions in quantum gravity, but much remains to be understood regarding the description of local bulk physics.
It is thus of interest to construct CFT observables which act like local bulk operators to the greatest degree possible.
This program is straightforward at leading  order in the $1/N$ expansion \cite{Witten:1998qj,Balasubramanian:1998sn,Banks:1998dd,Balasubramanian:1998de,Hamilton:2006az,Heemskerk:2012mn} where bulk locality can be exact.
But the situation becomes more complicated at higher orders in $1/N$ where all bulk fields interact with gravity.  One must then account for bulk gauge symmetries -- especially diffeomorphism-invariance --  which are associated with conserved currents in the boundary.
Fully local bulk operators then fail to exist (though see \cite{Marolf:2015jha}), and naively-local bulk operators require some form of ``dressing'' in order to be promoted to observables.

One approach to constructing such dressed operators is to simply fix a gauge in the bulk.
A popular choice for the metric is the Fefferman-Graham gauge $h_{z\mu} = 0$, which is analogous to the axial gauge $A_z = 0$ for a gauge field.  With this choice the leading $1/N$ corrections were computed in \cite{Kabat:2012hp,Kabat:2013wga,Kabat:2015swa}.   Since Fefferman-Graham coordinates are essentially Gaussian normal coordinates, the resulting observables may be described in a gauge-invariant way as the values of bulk fields at a certain (renormalized) distance into the bulk along one of the associated geodesics running inward from the AdS conformal boundary \cite{Heemskerk:2012np}.
They may thus be said to describe local operators dressed by a gravitational string. We follow \cite{Donnelly:2015hta} in referring to such observables as gravitational Wilson lines; see \cite{Tsamis:1989yu} for related work at zero cosmological constant.

Due to this gravitational string, acting with such operators modifies the gravitational field even if the leading-order bulk operator did not involve the graviton.  This is to be expected on general grounds, as the gravitational Gauss law requires any operator changing the total energy (momentum, angular momentum, etc) to alter the gravitational flux at infinity.  This fundamental observation dates back at least to the classic works of Arnowitt, Deser, and Misner \cite{Arnowitt:1959ah,Arnowitt:1960zzc,Arnowitt:1961zz,Arnowitt:1962hi}; see also \cite{Marolf:2008mf,Heemskerk:2012mn,Heemskerk:2012np,Almheiri:2014lwa,Giddings:2015lla,Donnelly:2015hta,Marolf:2015jha} for related comments and discussions of alternatives in a modern context.

The interesting feature of gravitational Wilson lines is that -- at the initial time at which they act -- the metric is modified only along the string.  So when acting on the vacuum such operators produce states in which the full gravitational flux emanating from a bulk source is initially tightly collimated along the associated geodesic.  As emphasized in \cite{Donnelly:2015hta}, this differs markedly from solutions like AdS-Schwarzschild, Kerr, etc. in which the gravitational flux is more uniform.
Analogous configurations in gauge theory, in which the electromagnetic field lines are highly collimated, have a long history; see e.g. \cite{Dirac1955,Bucholz1982,Steinmann1983,Steinmann2004}.

Though the string-like configurations for the gravitational field may be unfamiliar, the source determines only the total amount of gravitational flux and not its distribution.  The latter constitutes a new form of gravitational hair which the above observables simply `comb' into a highly concentrated initial state. The unfamiliar nature of the combed solutions is associated with the fact that they represent strong excitations above the familiar ones.  The concentration of flux raises the energy beyond the level that might otherwise be anticipated, and time evolution generally causes the string to spread out at the speed of light\footnote{Discussions of black hole hair typically focus on stationary solutions and so do not mention such configurations of dynamical flux.
But in the AdS$_3$ case on which we focus below, the Virasoro asymptotic symmetries \cite{Brown:1986nw} imply that our hairy excitations do not dissipate, but are instead periodic in time.
Interestingly, the asymptotically-flat limit of this hair gives precisely the degrees of freedom excited by the 2+1 analogue of Bondi-Metzner-Sachs transformations \cite{Barnich:2006av}.  While related hair has recently been suggested \cite{Strominger:2014pwa,Hawking:2015qqa,Dvali:2015rea,Hooft:2015jea} to be relevant to black hole information, in AdS$_3$ our excitations clearly provide only an $O(1)$ correction to the $O(\frac{\ell}{G})$ black hole entropy.}.  Indeed, one should expect the singular field produced by any exact Wilson line to have infinite energy, so that some transverse smearing will be required to produce finite-energy solutions.  The precise energy obtained will clearly depend on both the amount and character of smearing chosen.

Our purpose below is to study this energy/smearing relationship and other properties of combed solutions in an analytically-tractable context.  While we are motivated by a desire to understand observables in quantum gravity, we also wish to study effects associated with finite back-reaction on the geometry.
The lack of a sufficiently complete non-perturbative formulation of bulk quantum gravity\footnote{A study in the context of loop quantum gravity (see e.g. \cite{Thiemann:2007zz,Rovelli:2008zza,Gambini:2011zz}) might be possible, though to our knowledge no total energy operator (generating asymptotic time-translations) has been constructed within this approach.} thus limits us to the study of classical solutions. Even so, the lack of symmetries and non-linearities in the gravitational field equations suggest that the problem generally requires numerics.  But an exception occurs in 2+1 dimensions for pure Einstein-Hilbert gravity with cosmological constant where the lack of propagating degrees of freedom means that all solutions are (at least locally) given by empty AdS$_3$ for negative cosmological constant $\Lambda$, 2+1 Minkowski space for $\Lambda =0$, or dS$_3$ for $\Lambda >0$.
Once we have fixed any possible identifications, the different configurations of gravitational hair must be given by the action of diffeomorphisms.
These diffeomorphisms may be nontrivial at infinity, so different combings of the gravitational hair in 2+1 dimensions correspond to different ways of gluing the spacetime to its asymptotic boundary.

We focus on the AdS$_3$ case below, where the above diffeomorphisms are known as boundary gravitons.  Our bulk solutions will correspond both to conical defects (mass parameter $0 > M \ge -1$ in the conventions of \cite{Banados:1992wn}) and Ba\~nados-Teitelboim-Zanelli (BTZ) black holes ($M \ge 0$).  Rather than concern ourselves with the detailed relation to Wilson line observables, we simply study time-symmetric solutions in which the gravitational hair at the AdS$_3$ boundary is confined at $t=0$ to an angular interval $[-\alpha,\alpha]$; i.e., for which the boundary stress tensor takes precisely its vacuum value outside this interval. We find finite-energy solutions for each $\alpha \in [0,\pi]$. At linear order in the source strength $(M+1)$ their energy $E$ is independent of $\alpha$, but non-linearities force $E\to \infty$ as $\alpha \to 0$.  Interestingly, for each $M, \alpha$ the energy-minimizing solutions can be said to displace the source from its naively-expected location at the end of an associated (smeared) gravitational Wilson line.  For $M < - (1-\frac{\alpha}{\pi})^2 $ one may compensate for the above displacement by further increasing the energy, though this is impossible for conical defects with $M \ge - (1-\frac{\alpha}{\pi})^2 $.  We conclude that only in the former case can any (smeared) gravitational Wilson lines act as one might naively expect. In the black hole case $M \ge 0$ we find no preferred notion of zero-displacement solution.

Implications for $\Lambda =0$ can then be studied through an appropriate limit. We will not address the $\Lambda > 0$ case here, though dS$_3$ can be given useful notions of a boundary if treated as in \cite{Kelly:2012zc}.

We begin in section \ref{overview} with a more detailed description of the combed solutions, an outline of how they are to be constructed, and a discussion of the above-mentioned notion of displacement.  Results for AdS${}_3$ then appear in section \ref{explicit}. We close with some final comments in section \ref{disc}, including discussion of the $\Lambda=0$ limit.

\section{Framework and definitions}
\label{overview}

\begin{figure}
\includegraphics[width=6in]{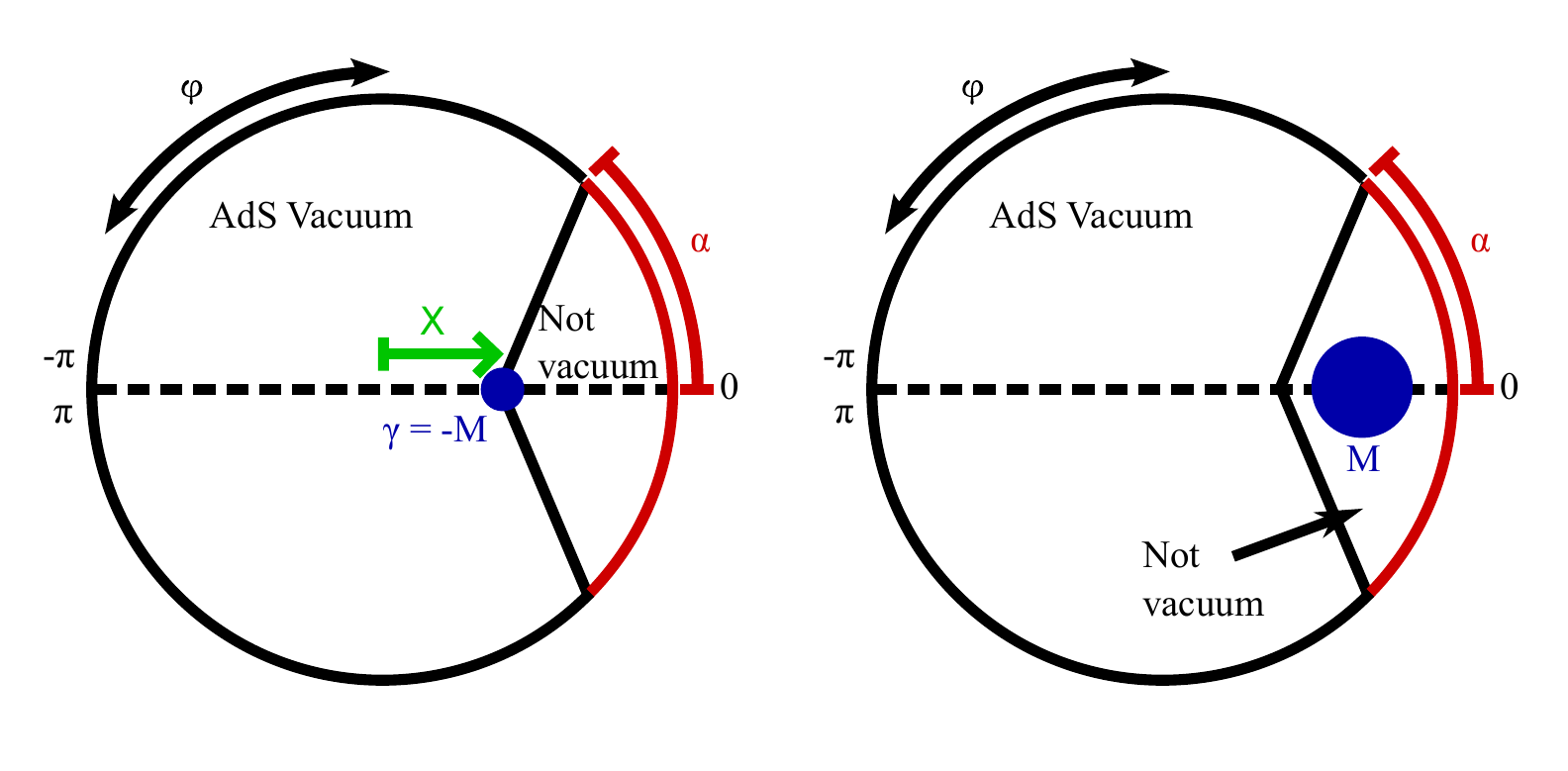}
\caption{Sketch of the desired solutions sourced by a conical defect (left) and a BTZ black hole (right).
We depict the $t = 0$ slice of the solution in coordinates such that the angular coordinate on the boundary is $\varphi$, and the coordinates in the vacuum region coincide with the Poincar\'e disk.
Since this sketch is only for illustrative purposes we leave the choice of coordinate in the dressed region unspecified.
Sources are indicated by disks (blue in color version).
The solutions are symmetric under $\varphi \to - \varphi$, with $\varphi$ the angle around the boundary circle.  The boundary stress energy tensor is that of the AdS vacuum everywhere except in the ``Not vacuum" region $\varphi \in [-\alpha, \alpha]$.  This allows us to interpret certain finite wedges of the bulk as pieces of a pre-existing vacuum preserved by the action of a smeared Wilson line.  For $M < 0$ we take the tip of the wedge to lie at the defect and use the shape of the resulting wedge to define an off-center displacement $X$.  For $M \ge 0$ we refrain from introducing a real-valued displacement $X$ due to the lack of a preferred vacuum wedge for black holes. }
\label{StateDiagram}
\end{figure}

We wish to study solutions of pure 2+1 Einstein-Hilbert gravity with negative cosmological constant corresponding to familiar bulk sources whose gravitational hair has been combed into unfamiliar configurations.  In general, gravitational hair may be characterized as those features of a gravitational solution that register far from the source. Some of this hair will be determined by sources.  Indeed, the lack of propagating bulk degrees of freedom in 2+1 dimensions means that for trivial topology the sources uniquely determine the resulting geometry.  Any further gravitational hair must thus be parametrized by diffeomorphisms acting on some reference solution.  But the only diffeomorphisms that are not gauge-redundancies are those that fall off slowly at the boundary \cite{Brown:1986nw}, and which are commonly called boundary gravitons. These are the degrees of freedom that define the hair of interest here. After briefly describing this structure in section \ref{SE}, section \ref{displace} introduces a useful measure of the degree to which the sources in such solutions have been displaced from the origin.  Throughout this work we will restrict to solutions satisfying both time-symmetry $t \to -t$ and a spatial reflection symmetry $\phi \to -\phi$.
The solutions of interest are depicted in figure \ref{StateDiagram}.

\subsection{Reference Solutions and Boundary Gravitons}
\label{SE}

We consider solutions in which, up to the above diffeomorphisms, the metric takes the form
\be
ds^2 = - \left ( \frac{r^2}{\ell^2} - M \right) dt^2 + \left (  \frac{r^2}{\ell^2} - M\right)^{-1} dr^2 + r^2 d\phi^2 \, ,
\label{globalMetricBH}
\ee
in terms of the AdS$_3$ scale $\ell$
with $\phi \in [0, 2 \pi)$ and $M \ge -1$. For $-1 \le M < 0$ the spacetime is global AdS$_3$ with a conical defect of deficit angle $\Delta \phi = 2\pi (1 - \sqrt{-M})$ inserted at the origin $r=0$.  For $M \ge 0$, the spacetime is a BTZ black hole.  We also require the conformal frame at infinity to be chosen so that the boundary metric is
\begin{equation}
\label{bmetric}
ds^2_\partial = g^{(0)}_{ab} dx^a dx^b = - \ell^{-2} dt^2 + d\phi^2.
\end{equation}
We will often display results in terms of the parameter $\gamma = -M$; this is especially useful for the conical defect case.

The set of all such bulk metrics is known explicitly up to gauge equivalence \cite{Banados:1998gg,Roberts:2012aq}, though we find it convenient to use the fact \cite{Brown:1986nw} that the gauge-equivalence classes of allowed diffeomorphisms are in one-to-one correspondence with the conformal isometries they induce on the boundary cylinder $S^1 \times {\mathbb R}$.   As in e.g. \cite{Kuns:2014zka}, using time-symmetry our diffeomorphism is in fact determined by the diffeomorphism $\phi \mapsto \varphi(\phi)$ that it induces on the boundary $S^1$ at the moment of time symmetry ($t=0$).

We may thus label all solutions of interest by the parameter $M$ and the function $\varphi(\phi)$, the latter may be said to describe the boundary gravitons to be added to the reference solution \eqref{globalMetricBH}.  One advantage of our framework is that our energy-minimizing solutions will fail to be continuous when expressed as in \cite{Banados:1998gg,Roberts:2012aq}; indeed, that form of the metric will contain Dirac delta-functions, as will the boundary stress tensor.  But $\varphi(\phi)$ will remain a continuous function, having discontinuities only in its derivatives.

As mentioned in the introduction, our notion of gravitational hair will be defined by the boundary stress tensor $T_{ab}$.  As is well-known from \cite{Balasubramanian:1998de}, the reference solutions \eqref{globalMetricBH} give
\begin{equation}
\label{refTab}
T_{tt} = \ell^{-2} T_{\phi \phi} = \frac{M}{16 \pi G\ell}, \ \ \ \text{with} \ \ \ T_{t\phi} =0.
\end{equation}
 The total energy is thus
\begin{equation}
\label{ETtt}
E = \int d \phi  \ \  n^\alpha \xi^\beta T_{\alpha \beta} = \frac{M}{8G}
\end{equation}
in terms of the future-pointing unit normal $n^\alpha \partial_\alpha = \ell \partial_t$ to the $t=0$ surface defined by the boundary metric \eqref{bmetric} and the time-translation Killing field $\xi^\alpha \partial_\alpha = \partial_t$.  The stress tensor $\tilde T_{ab}$ of the solution induced by $\phi \mapsto \varphi$ is readily computed using the fact \cite{Witten:1998qj,Balasubramanian:1998de,Henningson:1998gx,deHaro:2000xn}  that it transforms according to the standard anomalous transformation rule of 1+1 conformal field theories\footnote{See e.g.  \cite{DiFrancesco:1997nk}, though we use conventions where the $E$ of \eqref{ETtt} is minimized by the ground state. The conventions of \cite{DiFrancesco:1997nk} differ by a sign, so that the ground state maximizes \eqref{ETtt}.}
\be
\label{newTtt}
\tilde T_{tt} = \frac{1}{(\varphi^\prime)^2} \left ( T_{tt} + \frac{c}{12\pi \ell^2} \{ \varphi ; \phi \} \right ) \, 
\ee
in terms of the central charge $c = \frac{3 \ell}{2G}$,  where the second term is the Schwarzian derivative,
\be
\label{Schw}
\{ \varphi ; \phi \} =  \frac{\varphi^{\prime \prime \prime}}{\varphi^\prime} - \frac{3}{2} \left ( \frac{\varphi^{\prime \prime}}{\varphi^\prime} \right )^2  \, .
\ee
 Indeed, any property of the solutions can be computed from $\varphi(\phi)$.

We wish to construct solutions in which the addition of boundary gravitons combs all gravitational hair into the interval $[-\alpha, \alpha]$.  That is, for $\varphi \notin [-\alpha, \alpha]$ we require the boundary stress tensor $\tilde T_{ab}$ to agree with the vacuum values $T^{vac}_{ab}$ given by \eqref{refTab} with $M=-1$.  Since an energy-minimizing solution will be unique, it will respect the $\mathbb{Z}_2$ symmetry $\varphi \to -\varphi$.  So for simplicity we restrict attention to even solutions.  These are generated from \eqref{globalMetricBH} by odd diffeomorphisms $\varphi(\phi) = - \varphi(-\phi)$.

\subsection{Displacement from the origin}
\label{displace}

We are free to act on any of the above solutions with AdS$_3$ isometries.  For non-trivial functions $\varphi(\phi)$, most isometries break the time- or spatial-reflection symmetries and are not of interest.  But there remains a one-parameter subgroup given by AdS-translations along the axis defined by $\varphi = 0$ and $\varphi = \pi$, which we henceforth call the $X$-axis.

When acting on the reference solutions \eqref{globalMetricBH}, it is natural to speak of such translations as generating displacement of the source from the origin.  One would like to define a similar notion of displacement $X$ for general solutions so that we may minimize the energy holding constant both $X$ and $\alpha$.  The point here is that one might wish to have some notion of where the operators described in the introduction will act, and thus where the source will be created.

In particular, gravitational Wilson line operators cannot affect the $t=0$ geometry away from the gravitational strings they create.  Suppose we begin with the global AdS$_3$ vacuum given by \eqref{globalMetricBH} with $M=-1$ and act with a gravitational Wilson line running along the positive $X$ axis ($\phi =0$) and ending at the origin (where it will create a source).  Even after smearing the Wilson line over $\phi \in [-\alpha, \alpha]$ (say, by acting with rotations\footnote{Though we in principle allow very general notions of smearing, requiring only that the resulting operator be confined to the region $\phi \in [-\alpha, \alpha]$. }), the region $|\phi | \ge \alpha$ will remain precisely vacuum.  As a consequence, if the Wilson line creates a conical defect at its end, for all $|\phi | \ge \alpha$ we must find radial geodesics emanating from this defect that intersect the boundary orthogonally\footnote{In the sense defined by the given conformal frame, meaning that in Fefferman-Graham coordinates the geodesics remain at constant position along the boundary and move only in the radial coordinate (typically called $z$).}.  If instead the conical defect is created elsewhere, such geodesics will hit the boundary at angles characterized by a certain dipole pattern.  Matching this pattern to the one created by acting on \eqref{globalMetricBH}
with an AdS translation then allows us to assign a displacement $X$ from the origin to the conical defect created by our operator.  We emphasize that this $X$ is fundamentally defined by the region in which the operator does {\it not} act.  As a result, even though it is computed from the solution created by the operator, it measures the discrepancy between the location of our source and the location (by assumption, the origin) at which the smeared Wilson line acts on the original vacuum solution.  When acting on the vacuum, and so long as it creates a source at its endpoint, the supposed Wilson line must thus create a solution with $X=0$.

The BTZ case $M \ge 0$ does not have a well-defined source and appears not to admit a clean definition of displacement $X$; see further discussion in section \ref{disc}.   We will nevertheless define $X$ for this case by analytic continuation from the conical defect case $M < 0$. The resulting $X$ will be complex, though the difference $X_1 - X_2$ between any two $M \ge 0$ solutions will be real. The (constant) imaginary part serves as a reminder that for $M\ge 0$ solutions with combed gravitational hair we define no preferred zero-displacement solution.  In all cases, acting with a further AdS translation along the $X$ axis will simply shift $X$ as expected.  As a result, even for $M > 0$ the limiting regime in which $X$ becomes very large can still be said to represent displacement far from the origin.

\section{Solutions and Results}
\label{explicit}

We now discuss the explicit solutions, computing the energy $E$ as a function of the displacement $X$.  We first describe general results for the vacuum region and for the region with gravitational hair in sections \ref{VacReg} and \ref{Hreg}.  Section \ref{minoverX} minimizes $E$ with respect to $X$, while the final sections (\ref{lightheavy} and \ref{weakstrong}) describe various simplifying limits.

\subsection{Vacuum Region}
\label{VacReg}

We wish to investigate implications of the combing condition $\tilde T_{ab}= T^{vac}_{ab}$ for $\varphi > |\alpha|$.  Time-symmetry and tracelessness of $\tilde T_{ab}$ imply that it is sufficient to check the component $\tilde T_{tt}$ given by \eqref{newTtt}.  Through \eqref{Schw}, solving $\tilde T_{ab}= T^{vac}_{ab}$ amounts to integrating a third order differential equation and so will generally introduce three constants of integration.  These parameters correspond to the fact that $T^{vac}_{ab}$ is invariant under  AdS$_3$ isometries, of which the subgroup preserving time-symmetry is generated by AdS translations along our $X$-axis, translations along the orthogonal $Y$-axis, and rotations.  But since the latter two would generally break the symmetry $\varphi \to -\varphi$, in the vacuum region $|\varphi| > \alpha$  the allowed diffeomorphisms are labelled  by a single parameter related to our displacement $X$.

Since $\varphi(\phi)$ is odd and defines a diffeomorphism on the circle, it suffices to think of $\varphi$ as a diffeomorphism mapping the interval $[0,\pi]$ to itself.  In particular, $\varphi$ satisfies $\varphi(0)=0$ and $\varphi(\pi)=\pi$.  While the above-mentioned differential equation may be solved directly with such boundary conditions, it is instructive to construct the desired solutions by following a different path.   We begin with the conical defects $(M< 0)$.  In that case, though both are negative, the magnitude of $T^{vac}_{tt}$ is larger than that of $T_{tt}$ in the reference solution \eqref{globalMetricBH} by a factor of $\gamma^{-1} = -M^{-1}$.  On an interval $(\alpha, \pi]$   we may then set $\tilde T_{tt} = T^{vac}_{tt}$ by using the simple scaling
\be
\varphi = D_{\sqrt{\gamma}}(\phi) = \pi - \gamma^{-1/2} (\pi - \phi)
\ee
so long as $\pi - \alpha < \pi \sqrt{\gamma}$ (as required by positivity of $\varphi$ for $\phi \in (0,\pi]$).  We refer to the transformation $D$ as a dilation in the sense that this is the conformal transformation it would define if extended in a natural way to the planar covering space of our boundary cylinder, taking the origin of this plane to coincide with $\phi=\pi$.

As in section \ref{displace}, we are now free to apply an AdS translation $P_X$ by an amount $X$ along the symmetry axis without changing the value of $T^{vac}_{tt}$ near $\varphi=\pi$.  We take $X$ to be the Killing parameter along the associated Killing vector field of empty AdS$_3$ pointing toward $\phi =0$ (and thus away from $\phi = \pi$) and having unit norm at the origin.  As a result, $P_X$ acts on the boundary via the conformal isometry
\be
\label{PX}
P_X(\phi) = 2 \tan^{-1} \left(e^{-X/\ell} \tan \frac{\phi}{2} \right),
\ee
which also determines its effect on the combing parameter $\alpha$.

Indeed, for any $\gamma$ and every final combing parameter $\alpha$, the following algorithm gives a one-parameter family of functions $\varphi(\phi)$ labeled by a displacement $X$ and giving $\tilde T_{ab}= T^{vac}_{ab}$ for $\varphi > \alpha$.  We begin by defining the angle $\hat \alpha$ such that the action of the inverse translation $P_{-X}$ on the $t=0$ boundary maps the angular interval $[0, \alpha]$ to precisely $[0, \hat \alpha]$.  Then for $\pi - \hat \alpha < \pi \sqrt{\gamma}$ we may choose
\begin{equation}
\label{2steps}
\varphi_{X, \alpha} (\phi) = (P_X \circ D_{\sqrt{\gamma}})(\phi) \ \ \ \text{for} \ \ \ \phi > \phi_\alpha : = \varphi^{-1}(\alpha).
\end{equation}
This is the expected family of solutions to our problem.
In contrast, for $\pi - \hat \alpha > \pi \sqrt{\gamma}$ we will be unable to extend \eqref{2steps} to a diffeomorphism of the circle. For positive $X$, the action of $P_{-X}$ will expand the above interval so that $\hat \alpha > \alpha$.  So for $M < 0$ the allowed values of $X$ will range over some (open) half-infinite interval of the form $(X_0, \infty)$.  For $X=X_0$, the map \eqref{2steps} is just the inverse of the one that constructs the conical defect solution \eqref{globalMetricBH} by cutting a wedge out of the $M=-1$ vacuum solution; it does not represent a solution to our problem as it forbids us from satisfying the boundary condition $\varphi(0)=0$. By explicitly solving the relevant differential equation one may check that there are no further solutions for $M < 0$.

\begin{figure}
\centering
\includegraphics[width = 3in]{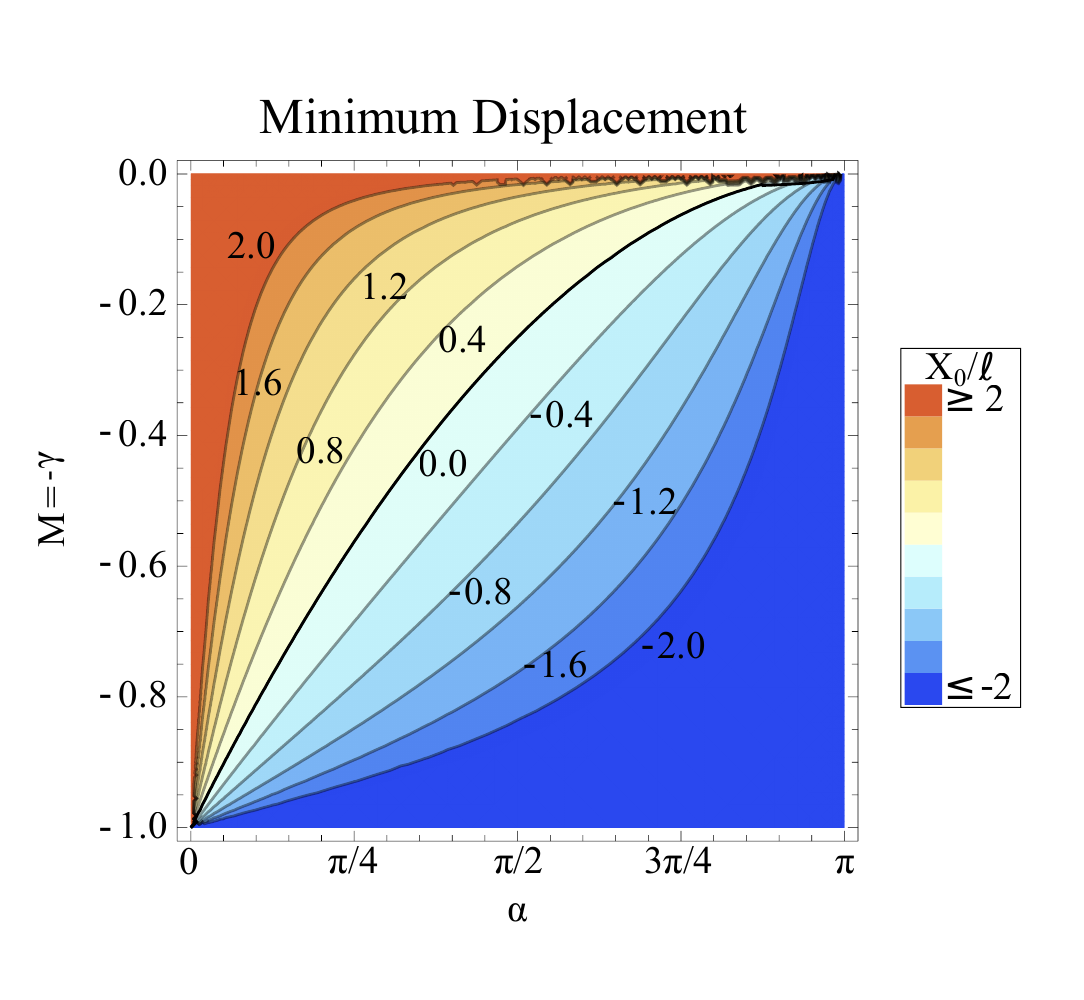}
\caption{A plot of the minimum displacement $X_0$ for conical defects against $\alpha$ and $M=-\gamma$.  The bold line shows $X_{0} = 0$, given by $\sqrt{-M} \pi = \pi-\alpha$.  Above this line all combed solutions suffer displacement in the direction of the gravitational hair.}
\label{MinDispPlot}
\end{figure}

Direct calculation gives the explicit form
\be
\varphi  = 2 \tan^{-1} \left [  e^{X/\ell} \tan (\sqrt{\gamma} (\phi - \pi) /2) \right ] + \pi \, ,
\label{vacTransform}
\ee
  and thus
\be
\label{ztova}
X= \ell \log \left [ \frac{\tan\left ( \frac{1}{2} ( \pi-\alpha)\right )}{\tan \left ( \frac{\sqrt{\gamma}}{2} ( \pi - \phi_\alpha ) \right )} \right ] \, .
\ee
Note that by using \eqref{ztova} we may equally-well label our solutions by $\phi_\alpha$ instead of $X$.
The value of $X_0$ is determined by setting $\phi_\alpha =0$ in \eqref{ztova}. In general, $X_0$ increases as we increase the defect mass or decrease the focusing angle, so that the defect is pulled toward the hairy region as we more tightly comb its gravitational hair. For fixed $\alpha$ we see that $X_0 \rightarrow -\infty$ (allowing all values of $X$) as $\gamma \to -1$ and $X_0 \rightarrow +\infty$ (allowing no finite values of $X$) as $\gamma \to 0$.
A contour plot of the minimum displacement $X_0$ as a function of $\alpha$ and $\gamma$ is shown in figure \ref{MinDispPlot}.

Note that setting $X=0$ (as for a gravitational Wilson line running to the origin, smeared without blurring the endpoint) requires
\be
\label{zeroX}
\phi_{\alpha, {X=0}} = \frac{1}{\sqrt{\gamma}} \left ( \sqrt{\gamma} \pi - \left ( \pi - \alpha \right ) \right ) \, .
\ee
Since $\phi_\alpha$ must be real and positive, we find that zero displacement can occur only if
\be
(\pi - \alpha) < \sqrt{\gamma} \pi \, .
\ee
This condition is illustrated by the bold line in figure \ref{MinDispPlot}.
In terms of the deficit angle $\Delta \phi$, the condition for existence of a zero displacement is $2 \alpha > \Delta \phi$; i.e. the angle subtended by the vacuum region must be less than the total angle around the defect.

Solutions for $M > 0$ may be found by analytic continuation of \eqref{vacTransform}.  The allowed values of $X$ are then complex, taking the form $X = - i \frac{\pi \ell}{2} + x$ for real $x \in (x_0, \infty)$. We find
\be
\label{x0}
x_0 = \ell \log \left [ \frac{\tan(\frac{1}{2} (\pi - \alpha))}{\tanh(\frac{\sqrt{M}}{2} \pi)} \right ] \, .
\ee
Our parametrization breaks down for $M$ precisely zero (where $x_0 \rightarrow \infty$), though one may take the $\gamma \to 0$ limit of \eqref{vacTransform} holding fixed $\sqrt{\gamma} e^{X/\ell}$.
As $M \to \infty$, \eqref{x0} simplifies  to $x_0 \to \ell \log  \tan \frac{\pi-\alpha}{2}.$
In all cases, applying a further AdS translation $P_{\Delta X}$ shifts $X \rightarrow X + \Delta X$ and acts in the natural way on $\alpha$.

\subsection{Hairy region}
\label{Hreg}

So long as it defines a diffeomorphism of the circle, any extension of \eqref{vacTransform} to $\phi < \phi_\alpha$ provides a solution with gravitational hair combed into the desired region $\varphi \in [-\alpha, \alpha]$.  But we are interested in solutions that minimize the energy for given $\gamma, \alpha, X$.
One expects the minimum energy to increase as $\alpha$ decreases, since a solution that is vacuum outside $[-\alpha,\alpha]$ is also vacuum outside of $[-\alpha', \alpha']$ when $\alpha' < \alpha$
\footnote{In particular, studying  orbits of the Virasoro group in detail (see e.g. \cite{Witten:1987ty,Garbarz:2014kaa}) shows that the minimum energy configuration within each orbit is rotationally symmetric.
This assures the physically sensible result that we cannot decrease the energy of a defect or black hole by focusing its gravitational dressing.}.

From \eqref{Schw} we see that if the first derivative of our transformation is discontinuous anywhere, the $(\varphi^{\prime \prime})^2$ term will give an infinite contribution to the energy.  We thus demand that $\varphi$ be $C^1$.  The energy of our final solution may be written $E = E_{\mathrm{vacuum \ region}} + E_{\mathrm{hair}}$ with $E_{\mathrm{vacuum \ region}} = - \frac{1}{8 \pi G} (\pi - \alpha)$ fixed by $\alpha$ and, since $\varphi(\phi)$ is odd,
\be
\label{hairE}
E_{\mathrm{hair}} = \frac{1}{8\pi G} \int_{0}^{\phi_\alpha}  \frac{1}{\varphi^\prime} \left [ - \gamma  + \left ( \frac{\varphi^{\prime \prime}}{\varphi^\prime}  \right )^2\right ] d\phi \, .
\ee
The form \eqref{hairE} was obtained from \eqref{ETtt}, \eqref{newTtt}, and \eqref{Schw} after integrating by parts to reduce the third derivative in \eqref{Schw} to a second derivative of $\varphi$.  It turns out that the associated boundary term exactly cancels the effect of an explicit delta-function contribution to $T_{tt}$ at $\varphi = \alpha$ associated with a possible discontinuity in $\varphi^{\prime \prime}$.

Minimizing \eqref{hairE} leads to a differential equation requiring $\tilde T_{tt}$ to be constant for $\phi \in (-\alpha, \alpha)$, with the value of this constant being chosen so that $\varphi$ is $C^1$.  The solution here may thus be constructed much as in the vacuum region, but with an extra dilation to scale the energy density to the desired value.  We write
\begin{equation}
\label{3steps}
\varphi_{X, \alpha} (\phi) = (\hat D_{\alpha/\hat \alpha} \circ P_{-\hat X} \circ \hat D_{\sqrt{\gamma}})(\phi) \ \ \ \text{for} \ \ \ \phi < \phi_\alpha : = \varphi^{-1}(\alpha).
\end{equation}
where the notation $\hat D$ indicates that these dilations are centered on $\phi =0$ (so $\hat D_{\sqrt{\gamma}}(\phi) = \gamma^{-1/2} \phi$ as opposed to the dilations $D$ of section \ref{VacReg} that leave fixed $\phi = \pi$), and the parameters $\hat \alpha, \hat X$ are fixed by requiring $\varphi$ to be $C^1$.  Explicit computation then gives
\begin{equation}
\label{HairExpSol}
\varphi_{X, \alpha} (\phi) =
 \frac{2\alpha}{\hat \alpha} \tan^{-1} \left [
\frac{\tan\left ( \hat \alpha/2 \right ) \tan (\sqrt{\gamma} \phi /2)}{\tan\left ( \sqrt{\gamma} \phi_\alpha/2 \right )}
  \right ] \, ,
 \ \ \ \text{for} \ \ \ \phi < \phi_\alpha : = \varphi^{-1}(\alpha),
\end{equation}
and

\be
1  = \frac{\sin \left ( \sqrt{\gamma}(\pi-\phi_\alpha)\right)}{\sin \left ( \sqrt{\gamma}\phi_\alpha\right)} \frac{\alpha }{\sin\left(\alpha\right)} \frac{\sin \left ( \hat \alpha\right )}{\hat \alpha} \, \ \ \
e^{\hat{X}/\ell} =  \frac{\tan\left ( \hat \alpha/2 \right )}{\tan\left ( \sqrt{\gamma} \phi_\alpha/2 \right )} \, .
\label{phiEqAppendix}
\ee
Unfortunately, the first equation in \eqref{phiEqAppendix} is non-algebraic and has no closed form solution for $\hat \alpha$.  Note that $\hat \alpha$ need not be real: $\varphi(\phi)$ will be real if $\hat \alpha$ is either purely real or purely imaginary, though more general complex $\hat \alpha$ are forbidden.   For $\varphi(\phi)$ to be monotonically increasing, $\hat \alpha$ must be positive real or positive imaginary.  It turns out that one can find such $\hat \alpha$ satisfying \eqref{phiEqAppendix} for any allowed choices of $\gamma$, $\alpha$, and $\phi_\alpha$.  In some cases there are multiple such $\hat \alpha$ (which will then all be real), though a more careful analysis of continuity for $\varphi$ (noting that the branch of $\tan^{-1}$ in \eqref{HairExpSol} is fixed by the condition $\varphi =0$ at $\phi=0$) shows that real $\hat \alpha$ must lie in the interval $\phi \in [0,\pi]$.  With this understanding there is a unique $\hat \alpha$ for each allowed triple $(\gamma, \alpha, \phi_\alpha)$.  The energy density of our solution satisfies
\be
T_{tt} =  - \frac{ 1}{16 \pi  G \ell} \frac{  \hat \alpha^2}{\alpha^2}  \ \ \ \text{for} \ \ \  |\varphi| < \alpha,
\ee
but due to the delta-function terms at $\varphi = \pm \alpha$ the total energy is
\bea
\label{FinalE}
E = && -  \frac{1}{8\pi G} \Big[ (\pi-\alpha) + \frac{\hat \alpha^2}{\alpha}  \nonumber \\ && + \frac{2}{ \sin(\alpha)} \left (\cos(\pi-\alpha) - \cos(\sqrt{\gamma}(\pi-\phi_\alpha))\right ) \nonumber \\ &&  - \frac{2}{\sin(\alpha)} \frac{\sin \left ( \sqrt{\gamma}(\pi-\phi_\alpha)\right)}{\sin \left ( \sqrt{\gamma}\phi_\alpha \right)}    \left ( \cos (\sqrt{\gamma} \phi_\alpha) - \cos(\hat \alpha) \right ) \Big] \, . \nonumber \\
\eea
In particular, for $M < 0$ we use \eqref{zeroX} to set $X=0$ and obtain
\be
\label{EzeroX}
E_{X=0} = - \frac{1}{8\pi G} \left [ (\pi-\alpha) + \frac{ \hat{\alpha}_{X=0}^2}{\alpha}  - 2 \frac{  \cos \left (  \pi \sqrt{\gamma} - (\pi - \alpha) \right ) -\cos( \hat{\alpha}_{X=0})  }{\sin  \left (  \pi \sqrt{\gamma} - (\pi - \alpha) \right ) } \right ]   \, ,
\ee
where $\hat{\alpha}_{X=0}$ obeys
\be
1  = \frac{\alpha}{\sin  \left (  \pi \sqrt{\gamma} - (\pi - \alpha) \right ) }  \frac{\sin \left ( \hat \alpha_{X=0}\right )}{\hat \alpha_{X=0}} \, .
\label{hatalphaXto0}
\ee

\begin{figure}
\centering
\includegraphics[width=3in]{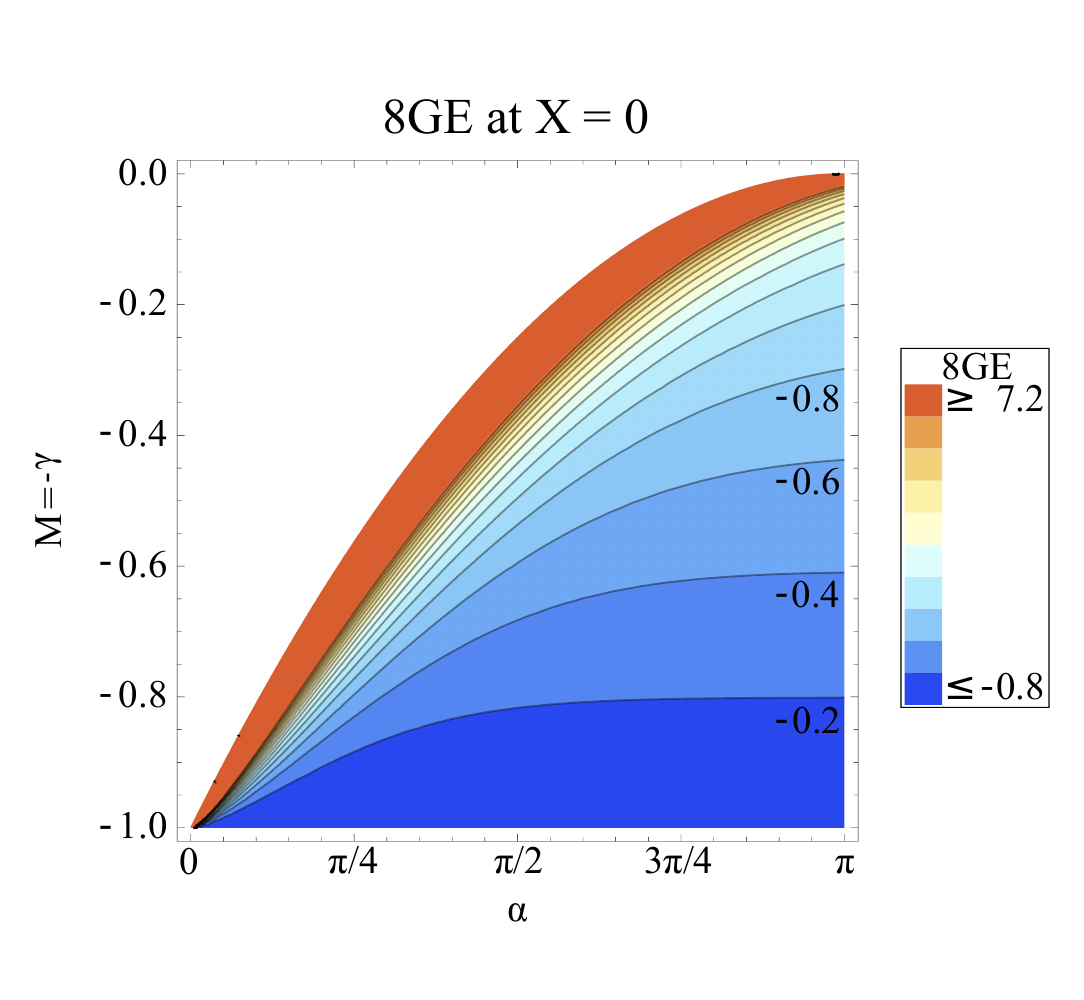}
\caption{   Minimum energies $E$ subject to the constraint $X=0$ for the conical defect regime $M \le 0$.  Note that $E \to \infty$ as $M \to 0$. Due to their increasing density, contours have been drawn only for $8GE \le 7.2$.  We define no preferred notion of zero displacement for black holes $(M \ge 0$).}
\label{zeroXFig}
\end{figure}

\begin{figure}
\centering
\includegraphics{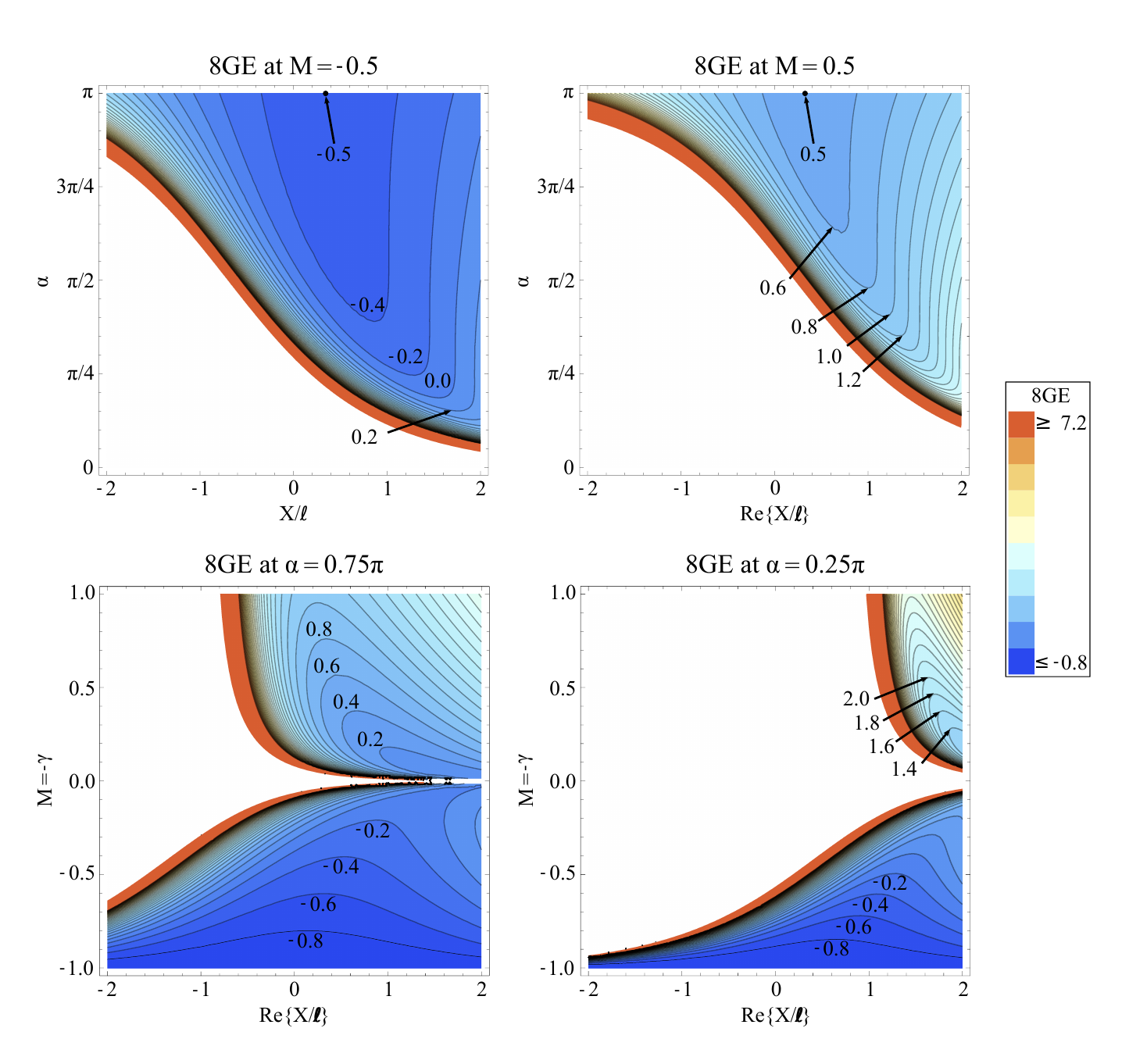}
\caption{Minimum energies with fixed mass $M$, combing parameter $\alpha$, and displacement $X$.  The white regions correspond to $X< X_0$ for conical defects or ${\rm{Re}} X < x_0$ for black holes; $E \to +\infty$ as such regions are approached.  Due to their increasing density, contours have been drawn only for $8GE \le 7.2$. The top row shows that, as discussed in section \ref{weakstrong} below, the the displacement $X_{\mathrm{min}}$ minimizing $E$ for given $\alpha, M$ is non-trivial even in the limit $\alpha \to \pi$. } \label{RelDisplacementPlot}
\end{figure}

While exact, the above expressions are not particularly enlightening.   We therefore plot numerical results with $X=0$  \eqref{EzeroX} below in figure \ref{zeroXFig} and for general $X$ \eqref{FinalE} in figure \ref{RelDisplacementPlot}.  The white in the latter (figure \ref{RelDisplacementPlot}) corresponds to the forbidden region $X \le X_0(M,\alpha)$; note that $E \to +\infty$ as $X \to X_0$.  This can be seen by noting that $\phi_\alpha \to 0$ (and thus $\hat \alpha \to \pi$ from \eqref{phiEqAppendix}) as $X \to X_0$.  The final line in \eqref{FinalE} thus dominates and gives a (divergent) positive contribution to $E$. Various simplifying limits will also be investigated in sections \ref{lightheavy} and \ref{weakstrong} below.

\subsection{Minimizing $E$ with respect to $X$}
\label{minoverX}

\begin{figure}
\centering
\includegraphics{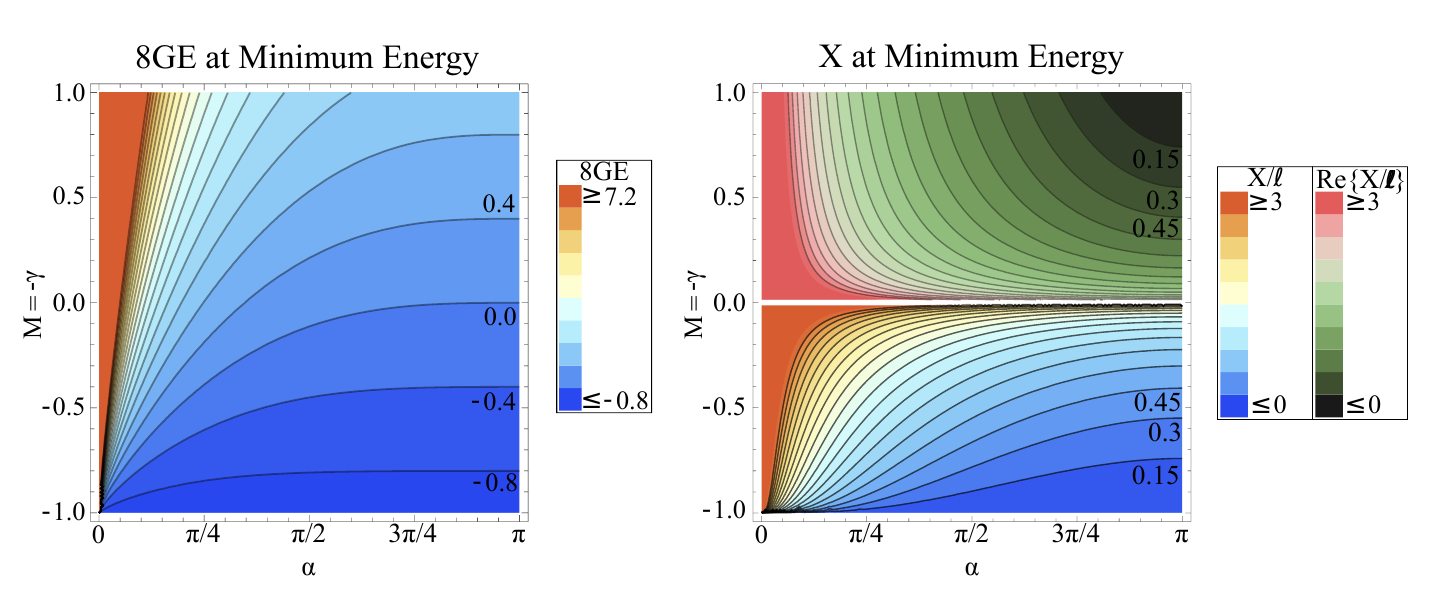}
\caption{The energy $E$ and displacement $X$ in the minimum energy configuration for a given $\gamma$ and $\alpha$.  Note that $X$ diverges at $\gamma=0$ and $X$ picks up a constant imaginary piece of $- \pi \ell/2$ at $\gamma >0$.  The minimum energy is smooth through this transition.  }
\label{MinEPlot}
\end{figure}

So far we have discussed minimizing $E$ while fixing $M = -\gamma$, $\alpha$, and the displacement $X$. But it is also interesting to find the displacement $X_{\mathrm{min}}$ that minimizes $E$ for fixed $M, \alpha$. It is technically simpler to do so by using $\phi_\alpha$ to parametrize the displacements.  We find the minimum to occur at

\be
\hat \alpha_{E_\mathrm{min}} = \sqrt{\gamma} \phi_{\alpha, E_\mathrm{min}} \, ,
\label{phiMin}
\ee
which requires
\be
\label{impza}
\frac{\sqrt{\gamma} \phi_{\alpha,E_\mathrm{min}}}{\alpha} = \frac{\sin \left ( \sqrt{\gamma}(\pi-\phi_{\alpha,E_\mathrm{min}})\right)}{\sin\left(\pi - \alpha\right)} \, ,
\ee
and so implicitly determines $\phi_{\alpha, E_\mathrm{min}}$ in terms of $\alpha$.  This
allows the minimum energy to be written
\be
E_{\mathrm{min}} =  - \frac{1}{8\pi G} \left[ (\pi-\alpha) + \frac{\gamma \phi_{\alpha,E_\mathrm{min}}^2}{ \alpha }   + \frac{2}{\sin (\alpha)} \bigg (  \cos (\pi - \alpha) - \cos (\sqrt{\gamma} (\pi - \phi_{\alpha,E_\mathrm{min}}))     \bigg ) \right]\, .
\label{minEEq}
\ee
The simplifications are associated with the fact that \eqref{phiEqAppendix} and  \eqref{phiMin} give $\hat X = 0$. The minimum energy configuration is thus the one for which the translation of the hairy region is trivial in \eqref{3steps}.  Thus full effect of \eqref{3steps} is then just a rescaling by $\alpha / \phi_\alpha$.  The minimum energy \eqref{minEEq} and the associated displacement $X_{E_\mathrm{min}}$ are plotted in Figure \ref{MinEPlot}.  Note that conical defects always have $X_{E_\mathrm{min}} > 0$, and $\mathrm{Re}(X_{E_\mathrm{min}}/\ell) >0$ for black holes.

\subsection{Light defects and heavy black holes}
\label{lightheavy}

Although the general formulae \eqref{FinalE}, \eqref{impza}, \eqref{minEEq} are quite complicated, one expects them to simplify in various limits.  One interesting choice is the limit of light conical defects (small deficit angle) $\epsilon_\gamma : = \gamma - 1 \rightarrow 0$ holding all other parameters fixed, in which the gravitational field is perturbatively close to the vacuum.  It is in this regime that we may attempt to compare with strictly perturbative treatments like \cite{Donnelly:2015hta}.  Here we will see that the combing affects the energy only at second order in $\epsilon_\gamma$.  In higher dimensions this would be clear from the fact that any measure of local energy density (e.g., the Landau-Lifshitz pseudo-tensor) in the gravitational field is quadratic in field perturbations and thus of order $\epsilon_\gamma^2$, and it turns out to be true here as well.  We also study the opposite limit $M \rightarrow \infty$.

The light defect limit is the only case where $E$ simplifies sufficiently that we may write an explicit closed form for general bulk displacement $X$.  To second order in $\epsilon_\gamma$ we find
\bea
E = && \frac{1}{8G} \Bigg[ -  1 + \cosh(X/\ell) \epsilon_\gamma  \nonumber \\ && + \frac{1}{4} \left ( \cosh(X/\ell) + \frac{\pi \left (  \cosh(2X/\ell) + \alpha \tan(\alpha) \sinh^2(X/\ell) - \sec(\alpha) \sinh(2X/\ell) \right )}{\tan(\alpha) -\alpha} \right ) \epsilon_\gamma^2 \nonumber
\\ &&
+ O(\epsilon_\gamma^3) \Bigg] \, .
\label{Elight}
\eea
The first term in \eqref{Elight} is just the energy of the displaced, unfocused insertion, and the combing-dependent correction is quadratic as claimed.  In this sense, the cost to comb the gravitational hair into any desired configuration is negligible for perturbatively light sources.  Note, however, that this is true only for $\epsilon_\gamma \to 0$ with $\alpha$ and $X$ fixed.  Taking $\alpha \to 0$ for fixed $\epsilon_\gamma$ and $X$ causes the quadratic term to diverge as $\alpha^{-3}$ so that it is no longer subleading.  Comparing the linear and quadratic terms suggests\footnote{A different conclusion would be reached by comparing either the linear or quadratic term with the constant term.  But the constant term is just the energy of the vacuum, which is related to the overall choice of zero-point for $E$.  It is the discrepancies from this value that are of physical interest.} that the expansion \eqref{Elight} is useful for $\epsilon_\gamma \ll \alpha^3$.

We may easily determine the zero-displacement energy by setting $X=0$ to obtain
\be
\label{EzeroXlight}
E_{X=0} = \frac{1}{8G} \left[ -  1 +  \epsilon_\gamma + \frac{1}{4} \frac{  \pi - \alpha + \tan (\alpha)}{ \tan(\alpha) - \alpha} \epsilon_\gamma^2 + O(\epsilon_\gamma^3)  \right] \, .
\ee
To instead choose $X$ to minimize the energy, one must satisfy \eqref{phiMin} which yields
\be
\frac{X_{E_\mathrm{min}}}{\ell} = \frac{\pi}{2 }  \frac{1 }{ \sin(\alpha) -\alpha \cos (\alpha)}  \epsilon_\gamma + O(\epsilon_\gamma^2) \, .
\ee
A displacement of order $\epsilon_\gamma$ from $X_{E_\mathrm{min}}$ (such as the displacement to $X=0$) provides a correction to the energy at $O(\epsilon_\gamma^3)$, and so to quadratic order the minimum energy matches the zero displacement energy:
\be
\label{Eminlight}
E_{X=0} = E_{\mathrm{min}}  + O(\epsilon_\gamma^3) \, .
\ee

We cannot impose $X=0$ in the complementary limit $M = -\gamma \to \infty$.  But choosing $X$ to minimize the energy gives
\be
E_{\mathrm{min}} =  \frac{1}{8G} \left[ \frac{ \pi M}{ \alpha}  + \frac{2\sqrt{M}}{ \alpha } \left ( 1 -  \log \left ( \frac{2\pi  \sin(\alpha)}{\alpha} \sqrt{M} \right ) \right ) \right] \, ,
\ee
\be
\frac{X_{E_\mathrm{min}}}{\ell} = - \frac{i \pi }{2}  + \log \left (\tan \left (\frac{1}{2} \left ( \pi - \alpha \right ) \right ) \right ) + \frac{\alpha}{\sin(\alpha)} \frac{1}{\pi \sqrt{M}}  + O(M^{-1})
\ee
Note that to leading order in $M$ the energy is just $\frac{\pi}{\alpha}$ times the energy of the uncombed solution $\eqref{globalMetricBH}$.

\subsection{Weak and strong combing}
\label{weakstrong}

Other interesting limits to consider are $\alpha \to \pi$, where the combing can be very weak, and the strong-combing limit $\alpha \to 0$.  In the former case the minimum-energy configuration must approach the reference solution \eqref{globalMetricBH}.     Indeed, introducing $\epsilon_\alpha = \pi - \alpha$ and choosing $X$ to minimize $E$ (i.e., to satisfy \eqref{phiMin})  yields
\be
E_{\mathrm{min}} = \frac{1}{8G} \left[ M  + \frac{(1+M)^2}{12 \pi} \epsilon_\alpha^3 + O(\epsilon_\alpha^5) \right] \, .
\label{AlphaToPi}
\ee
As expected, the first term is precisely the energy of \eqref{globalMetricBH}.  Interestingly, the corresponding displacement $X$ is
\be
\label{chiminsma}
\frac{X_{E_\mathrm{min}}}{\ell} = - \log {\sqrt{-M }} + \frac{ (1+M)}{4} \epsilon_\alpha^2 + O(\epsilon_\alpha^4) \, ,
\ee
which approaches the non-zero constant $- \log \sqrt{-M}$ as $\epsilon_\alpha \to 0$.  While surprising at first sight, there is nothing inconsistent about this result; it reflects the fact that the displacement is defined using the vacuum region which disappears (and thus degenerates) in the desired limit.

Recall that $\alpha \to \pi$ implies  $X_0 \to -\infty$ for conical defects, allowing us to fix any real value of $X$.  And for $M > 0$ black holes it yields $x_0 \to -\infty$, allowing any choice of ${\mathrm{Re}(X/\ell)}$.  In general we find
\bea
\label{apigenchi}
E = - \frac{1}{8 \pi G} \Bigg [ && \frac{\hat{\alpha}_\pi^2}{\pi} +2 \frac{ e^{-X/\ell} \left ( \cos(\hat{\alpha}_\pi)  - \cos(\sqrt{\gamma} \pi)\right)}{\sin(\sqrt{\gamma} \pi)}  \nonumber \\ && - \left (\frac{ e^{-X/\ell}  \left ( \cos(\hat{\alpha}_\pi)  - \cos(\sqrt{\gamma} \pi)\right)}{\sin(\sqrt{\gamma} \pi)} \right )^2\epsilon_\alpha + O(\epsilon_\alpha^2) \Bigg ] \, ,
\eea
where $\hat \alpha_\pi$ is the value of $\hat{\alpha}$ at precisely $\alpha = \pi$; i.e. $\alpha_\pi$ satisfies
\be
1  = \frac{\pi e^{-X/\ell}}{\sin(\sqrt{\gamma} \pi)} \frac{\sin (\hat \alpha_\pi)}{\hat \alpha_\pi} \, .
\ee
In particular, as suggested by \eqref{chiminsma}, setting $X=0$ does not give the energy of our reference solution.  Instead, the case $X=0$ is an excitation above the reference solution by an amount that cannot be neglected even in the limit $\alpha \to \pi$.

\begin{figure}
\centering
\includegraphics[scale=1]{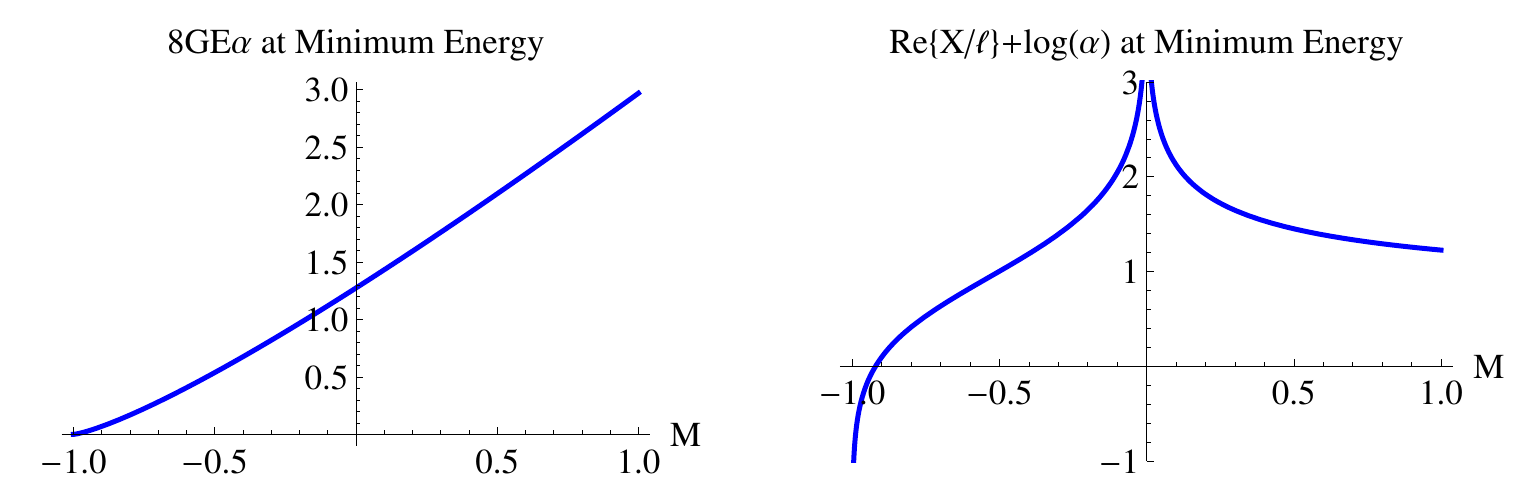}
\caption{ Plots of $8GE\alpha $ and $\mathrm{Re}\{X/\ell\}+\log(\alpha)$, with $X$ chosen to minimize $E$ for each $M$, in the limit $\alpha \to 0$; i.e., the plots show the coefficients of the leading terms in \eqref{Esma} and \eqref{chisma}.  Recall that $X$ is real for $M<0$, but has a constant $-i \pi \ell/2$ term for $M>0$. }
\label{AtoZeroZa}
\end{figure}

The complementary limit $\alpha \to 0$ localizes the hair near $\varphi=0$.  The choice $X=0$ is forbidden for any fixed $M > -1$, but we may take $X$ to minimize the energy by imposing \eqref{impza}.
Setting $\alpha =0$ in \eqref{impza} defines a parameter $\phi_{0, E_\mathrm{min}}$ that satisfies
\be
\sqrt{\gamma} \phi_{0, E_\mathrm{min}} = \sin \left ( \sqrt{\gamma}(\pi-\phi_{0, E_\mathrm{min}})\right) \, .
\label{AtoZeroZaEq}
\ee
The expansions of the minimum energy $E_{\mathrm{min}}$ and associated displacement $X_{E_{\mathrm{min}}}$ for small non-zero $\alpha$ are naturally written in terms of this parameter and take the forms
\bea
\label{Esma}
E_{\mathrm{min}} =  \frac{1}{8G} \bigg[  && \frac{1}{\pi \alpha} \left ( - \gamma \phi_{0, E_\mathrm{min}}^2 + 2 \cos(\sqrt{\gamma}(\pi - \phi_{0, E_\mathrm{min}})) + 2  \right ) -1  \nonumber \\ &&  + \frac{1 }{3 \pi} \left ( 1 + \cos (\sqrt{\gamma}(\pi - \phi_{0, E_\mathrm{min}})) \right ) \alpha + O(\alpha^2) \bigg]
\eea
and
\be
\label{chisma}
\frac{X_{E_\mathrm{min}}}{\ell} = - \log \alpha + \log \left (2 \cot \left (\frac{\sqrt{\gamma}}{2}(\pi - \phi_{0, E_\mathrm{min}})\right ) \right )  \\
 +  \frac{1}{12} \tan^2 \left (\frac{\sqrt{\gamma}}{2}(\pi - \phi_{0, E_\mathrm{min}}) \right ) \alpha^2 + O(\alpha^3) \, .
\ee
The leading terms in the energy is proportional to $1/\alpha$, while the displacement blows up as $- \log \alpha$.  The coefficients are complicated, so we plot them in figure \ref{AtoZeroZa}. Ignoring the vacuum (constant) term in \eqref{Esma} and comparing the terms of order $\alpha^{-1}, \alpha$ suggests that the expansion is valid for $\alpha^3 \ll \epsilon_\gamma$, which is precisely complementary to the regime required for \eqref{Elight}.    Although it diverges for small $\alpha$, we note that the energy remains close to the vacuum value for $\alpha^3 \gg \epsilon_\gamma^4$.

\section{Discussion}

\label{disc}

Our work above studied the combing of gravitational hair from standard sources in AdS$_3$.  The sources considered were conical defects $(-1 \le M < 0)$ and BTZ black holes $(M \ge 0$), both with zero angular momentum.  We showed explicitly how the gravitational flux from such sources can be combed into an angular interval $\varphi \in [-\alpha, \alpha]$ for any $0 \le \alpha \le \pi$ as measured at infinity.  However, this combing is generally associated with some displacement $X$ from the origin -- a notion which we defined sharply only for the conical defect case $M = -\gamma < 0$.
Each of our solutions also has an associated displacement $\hat X$ as measured from the the hairy region.
It would be interesting to understand if $\hat X$ is related to the 3+1 perturbative commutator found in \cite{Donnelly:2015hta}, which takes the form of a constant-magnitude radial displacement throughout the hairy region.

For perturbatively light defects $(M \to -1)$  the displacement $X$ can be chosen arbitrarily and the minimum energy configurations can be studied as functions of the triple $M, \alpha, X$.  Consistent with perturbative expectations in higher dimensions (and in particular with the results of \cite{Donnelly:2015hta}), we found combing to affect the energy only at second order in the source strength $(1+M)$.  But for finite-strength sources our $X$ must diverge as $\alpha \to 0$.  Indeed, zero-displacement solutions require
\begin{equation}
\label{OK}
(\pi - \alpha) < \sqrt{\gamma} \pi;
  \end{equation}
i.e., that the conical deficit is smaller than the angular size of the hairy region.

Our work was motivated by the study of gravitational Wilson line observables.  We noted that one expects such observables to create sources at their endpoints, but -- even when appropriately smeared -- to preserve a wedge-shaped region of the vacuum.  Though such properties may be verified perturbatively, any extrapolation to non-perturbative quantum gravity is at best a conjecture.  Indeed, the non-perturbative theory may admit multiple equally-natural extensions of the perturbative and semi-classical definitions of such an observable, each associated with different properties. Our results clearly constrain such extensions, suggesting that operators one might wish to call gravitational Wilson lines ending at the origin can create sources at their endpoints only in the regime \eqref{OK}.    In particular, the notion of an unsmeared Wilson line ($\alpha =0$) can be sensible only at the perturbative level.

There may, however, be a notion of an operator $O$ that creates a source in the expected location while producing gravitational hair in a region ${\cal R}_O$ of some shape more general than the wedges defined by angular intervals $[-\alpha, \alpha]$ that we have discussed thus far.  The `location' at which $O$ acts would then be defined by choosing a region ${\cal R}^{vac}_O$ of the AdS$_3$ vacuum to be left invariant (and thus free of gravitational hair) by the action of $O$.  Since there is no local notion of gravitational hair in 2+1, the above requirement simply means that $O$ creates a solution diffeomorphic to some reference solution (say, of the form \eqref{globalMetricBH}) whose $t=0$ initial data surface may be cut into two pieces: ${\cal R}_O$, which is a specified region of the same reference solution, and ${\cal R}^{vac}_O$  which is a specified region of the AdS$_3$ vacuum. Both pieces come labelled with a preferred angular coordinate $\varphi$ such that, when sewn together, $\varphi$ defines a continuous map from $S^1$ to the $t=0$ boundary.

Given a fixed reference solution, ${\cal R}^{vac}_O$ is determined by ${\cal R}_O$  up to diffeomorphisms; the geometry of ${\cal R}_O$ cannot be chosen independently.  For conical defects ($M< 0$), we may take
${\cal R}_O$ to be any connected region of the $t=0$ surface in \eqref{globalMetricBH} that includes some interval along the boundary.

Our notion of displacement $X$ was sharply defined only for conical defects, as it is only for conical defects that the source is truly localized.  In contrast, creating a BTZ black hole necessarily requires an operator to act over a finite-sized region that includes the black hole.  In particular, one should be able to proceed as in the above paragraph, though in the BTZ case the allowed regions ${\cal R}_O$ will be more complicated to describe.  It may nevertheless be interesting to characterize the possible such constructions and in particular the corresponding intervals $[-\alpha, \alpha]$ of $\varphi$ along the boundary into which they allow gravitational flux to be combed.  It would also be interesting to study possible connections with \cite{Almheiri:2014lwa,Pastawski:2015qua}, which suggested that commutators with operators creating black-hole-like objects should be rather different than with the perturbative gravitational Wilson lines.  The conjecture that naively-acting operators can exist only above some minimal value of $\alpha$ is very much along these lines.

We have so far focussed on the case of negative cosmological constant and AdS$_3$ length scale $\ell$.  But one may also wish to study solutions with zero cosmological constant.  Since a cosmological constant is irrelevant on short distance scales, all such solutions may be obtained from those studied above by taking $\ell \to \infty$ holding all other parameters fixed.

The lack of 2+1 asymptotically flat black holes suggests that we limit discussion to the conical defect case $\gamma = -M > 0$.  For finite $\ell$, this setting provided a clean notion of the absolute displacement $X$ of the combed solution from the center of the spacetime.  To have a well-defined asymptotically flat limit, $X$ should remain finite as $\ell \to \infty$.  Thus we require $X \to 0$.  So combed solutions with zero cosmological constant exist only for $(\pi - \alpha) < \sqrt{\gamma} \pi$.

At least up to what may be regarded as a conventional shift to set the energy of 2+1 Minkowski space to zero, or to $-\frac{1}{4G}$ as in \cite{Marolf:2006xj},
the energies of such solutions continue to be given by \eqref{EzeroX}; see also the limiting results \eqref{EzeroXlight} and \eqref{apigenchi} for $X=0$.  Due to our normalization of the time-translation Killing field, these energies are already independent of $\ell$.  Once the asymptotically flat limit has been taken, we are of course free to act with any additional asymptotically flat translation without further changing the energy. Interestingly, such combed solutions do not satisfy the standard definition \cite{Ashtekar:1993ds} of asymptotically flat 2+1 spacetimes which turns out to require the gravitational flux to become rotationally symmetric at infinity.  This suggests (as do the results of \cite{Barnich:2006av}) that it will be useful to generalize the definition of \cite{Ashtekar:1993ds}, though we leave this for future work.

Thus far we have restricted attention to solutions that are vacuum outside the location of some given source.  We have thought of the source as associated with some `bare' operator to which some gravitational dressing must be applied.  One may allow the dressing to add further matter excitations as well, though since positive energy matter will largely act to strengthen the source  this seems unlikely to allow interesting exceptions to the constraints found above. The point is instead that beyond 2+1 dimensions there is no sharp distinction between a purely gravitational dressing and one that involves matter, as quantum pair creation processes generically require the former to evolve into the latter.

It remains to discuss further implications for higher dimensions.
At least at the classical level, in this context point particles can exist only perturbatively in $G$.  A non-perturbative analysis must thus consider operators that create sources over finite regions.  Their action on the vacuum will again be naturally described by choosing a region ${\cal R}_O$ over which the operator is to act and a region ${\cal R}^{vac}_O$ of the vacuum to be left invariant.  Now, however, the possible addition of gravitational waves allows more freedom in choosing the geometry inside ${\cal R}_O$.   Even in the black hole regime it thus seems likely that, as measured at infinity, we will be able to freely specify the cone of solid angle into which the gravitational hair may be combed.  Indeed, as we found in 2+1 dimensions, one would generally expect $E \to +\infty$ as one approaches the boundary of any forbidden region.  And aside from singular limits analogous to our $\alpha =0$ the bulk geometries should remain smooth.  So such infinite energies must be associated with IR divergences, which necessarily weaken as the dimension is increased.

\section*{Acknowledgements}
We thank Veronika Hubeny and the KITP program ``Quantum Gravity Foundations: UV to IR'' for stimulating us to undertake this project.  We also thank Veronika for collaboration during the initial stages of the project and acknowledge useful discussions with Eric Dzienkowski, Sebastian Fischetti, Steve Giddings, and Kevin Kuns.   E.M. was supported by the National Science Foundation under grant number NSF PHY13-16748.  D.M. was supported by the National Science Foundation under grant numbers PHY12-05500 and PHY15-04541.  He also thanks the KITP for their hospitality during the initial stages of the project, where his work was further supported in part by National Science foundation grant number PHY11-25915. This work was also supported by funds from the University of California.

\bibliographystyle{JHEP}
\bibliography{DressingBib}

  \end{document}